# TWO PROTO-SCIENCE-FICTION NOVELS WRITTEN IN FRENCH BY EIGHTEENTH CENTURY WOMEN

## Yaël Nazé*

*University of Liège, B5C, Allée du 6 Août 19c, B4000-Liège, Belgium.*
E-mail: ynaze@uliege.be

**Abstract:** With Cyrano, Voltaire, and Verne, France provided important milestones in the history of early science fiction. However, even if the genre was not very common a few centuries ago, there were numerous additional contributions by French-speaking writers. In this paper, we review two cases of interplanetary novels written in the second half of the eighteenth century and sharing a rare particularity: their authors were female. *Voyages de Milord Céton* was imagined by Marie-Anne de Roumier-Robert whereas Cornélie Wouters de Wasse conceived *Le Char Volant*. While their personal lives were very different, and their writing style too, both authors share in these novels a common philosophy in which equality—between ranks but also between genders—takes an important place. Their works thus clearly fit into the context of the Enlightenment.

## 1 INTRODUCTION

Early science fiction may be traced back to antiquity, but the roots of its modern avatar are often attributed to the Age of Reason. Of course, there is no single country, language, or civilization that may claim to be the unique point of origin. This contribution will focus on the case of France. The country has produced several influential works in this context. In 1657 Savinien de Cyrano de Bergerac (1619–1655, Figure 1) proposed a comical journey to the lunar states and empires (soon to be complemented by a journey to their solar colleagues). The lively text, with its satirical hue, was inspired by the novel *The Man in the Moone* written by Francis Godwin (1562–1633) and translated into French in 1648 by Jean Baudouin (1590–1650). Cyrano's novel was a great success, influencing many other works there-after. Two centuries later, Jules Verne (1828–1905, Figure 1) proposed another diptych, *De la Terre à la Lune* (1865) and *Autour de la Lune* (1869), this time limited to a journey to the Moon. Verne's tone was very different from Cyrano's: no satire, no humor, but detailed descriptions of the technology and knowledge of the time. These first *hard* science-fiction novels were also immensely influential, well beyond France. Indeed, many astronautics pioneers (notably Konstantin Tsiolkovsky (1857–1935), Robert H. Goddard (1882–1945), and Wernher von Braun (1912–1977)) quoted Verne's books as inspiring sources.

Cyrano and Verne are bookends of sorts to the eighteenth century, famous for its Enlightenment. Often neglected in this context, the epoch actually experienced a science-fiction boom, with many works published on journeys of various types. One often presents *Micro-*

*mégas*, by Voltaire (1694−1778), published in 1752, as the main science-fiction story of this time. In it, a big (really big) extraterrestrial from Sirius travels to the Solar System where he meets a much smaller (but still big by our standards) creature from Saturn. Together, they further travel to the Earth, at first thinking it uninhabited then realizing it harbors microscopic people. While typically Voltairian, this short text mocking the importance that the earthlings attach to themselves somehow recalls Cyrano's writings because of its satirical tone. In addition to these famous stories, there were many other

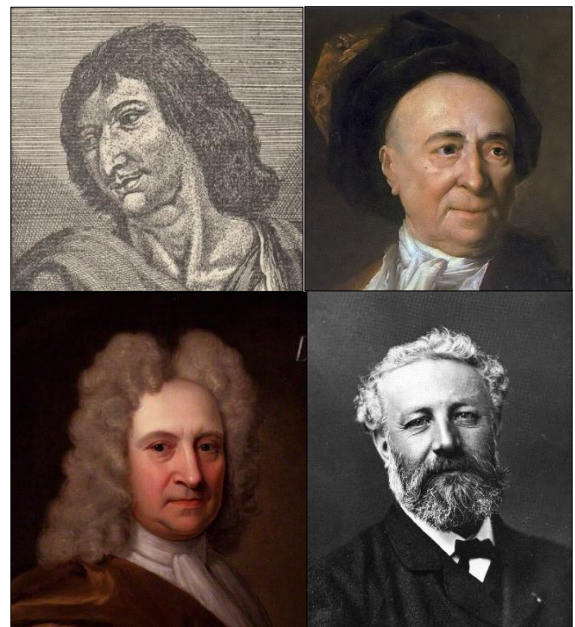

Figure 1: While no portrait of de Roumier-Robert or Wouters is known, the situation appears different for other protagonists of this article. From top left to bottom right are shown Cyrano de Bergerac (engraving, from gallica.bnf.fr / BnF - ark:/12148/bpt6k6206781d), Fontenelle (painting by N. de Largillière, from Wikipedia – public domain), Edmund Halley (portrait by R. Phillips, before 1722, from National Portrait Gallery #4393), and Jules Verne (photograph by F. Nadar around 1878, from Wikipedia – public domain).

* FNRS Senior Research Associate

works presenting interplanetary journeys. Although less influential, they are no less interesting. This paper focuses on two works, the *Voyages de Milord Céton* and *Le Char Volant*, which share a peculiarity: they were written by women. The next two sections will examine the two novels, each one first presenting the author, then the plot, and finally the work's reception and its inspiration sources. Section 4 will further study the astronomical and philosophical content of the novels, while Section 5 summarizes and concludes.

## 2   VOYAGES DE MILORD CÉTON

### 2.1   The Author

Marie-Anne de Roumier-Robert was born in 1705 to a wealthy French family, bourgeois but probably of noble origins (Harth, 1992). Her father was a friend of Bernard Le Bouyer de Fontenelle[1] (1657–1757, Figure 1). As a young girl, she may have heard at home some scientific discourse by that famous scholar (La Porte and La Croix, 1767: 79). The family was however ruined in the aftermath of the John Law scandal[2] in 1720 and her parents died rather quickly afterwards. As a consequence, Marie-Anne's education had to be stopped and she was sent to a convent before finally being married to a lawyer named Mr Robert. Despite her isolation and poor health, she wrote several novels before her death in 1771 (La Porte and La Croix, 1769, editor's introduction to de Roumier-Robert, 1787a, and Harth, 1992). The one we examine in this paper, *Voyages de Milord Céton dans les Sept Planettes ou Le Nouveau Mentor*, was first published in 1765–1766 (Figure 2).

### 2.2   The Novel

The title page of the novel mentions "Mrs R.-R." as the translator. This peculiarity is explained in the preface through the following story. One day, a small 'salamander' jumped out of the fire and discussed with her. It was maybe just a dream but, after that interlude, she found sheets of paper on her desk full of text (in her own handwriting!). She then explained that the following pages simply reproduce that text and as such she merely was an editor or a translator, not the real author (although she actually was, of course!). This stratagem was quite common in the past, especially in fantasy or science-fiction novels, but here it further served the purpose of making a woman acceptable as a source of literature.

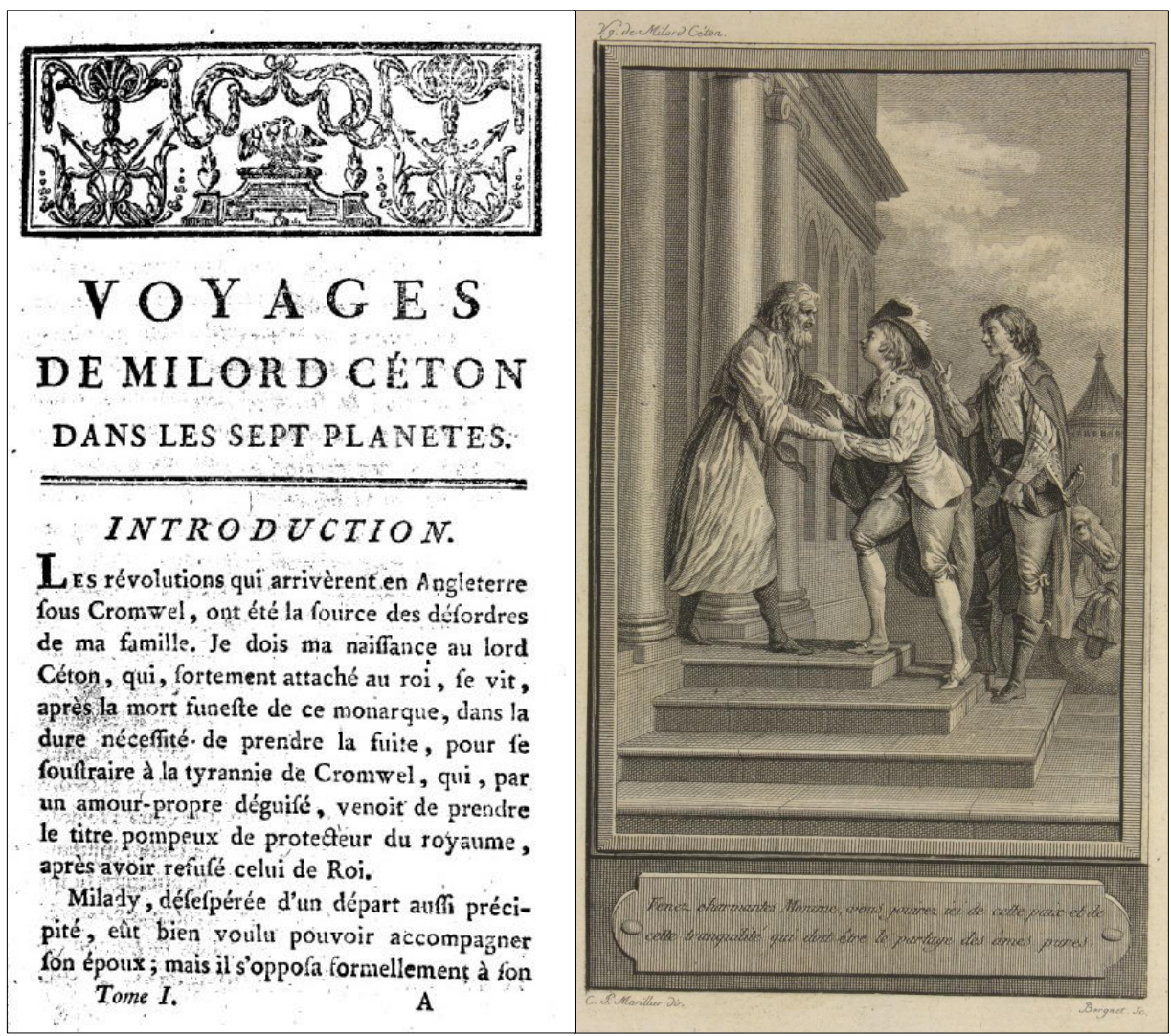

Figure 2: Front page of *Voyages de Milord Céton* by de Roumier-Robert (1787a, from gallica.bnf.fr/BnF - ark:/12148/bpt6k818002) and first of the four engravings made by C.-P. Marillier (1740–1808) to accompany the text in that edition (from Bibliothèques Universitaires de Poitiers, Fonds ancien, FD 1511). This engraving relates to the very beginning of the story and shows the genie, pictured as an old, wise man, welcoming Monime (in man's clothes) with Céton behind her.

The novel then begins with an introduction. The narrator is an English orphan boy, Céton, whose father had to flee after Cromwell's revolution and whose mother died from sorrow soon afterwards. His sister Monime and he were hosted by a distant relative when they stumbled upon Zachiel, a genie, who decided to provide them with a good education (Figure 2). Towards this aim, Zachiel envisaged a space voyage, which Céton soon accepted, pushed by Monime's enthusiasm.

The novel then continues with seven chapters, one for each visited 'planet': Moon, Mercury, Venus, Mars, Sun, Jupiter and Saturn, by order of appearance. Except for the Sun and Saturn, the 'planets' actually constitute opportunities to show the worst human flaws: the Moon has inconstant (lunatic?) and frivolous inhabitants; avarice and greed rule over Mercury; short-lived passion and libertinism are found on Venus; Mars welcomes useless war and unwonted glory; while Jupiter welcomes plotting courtiers of all sorts because nobility reigns on this planet, even if titles were unfairly acquired. In each case, the plot structure is similar. After a depiction of the planetary vices in various situations, the two orphans encounter a lonely virtuous soul who has been persecuted and whom they help. In contrast, they find the Sun to be the seat of reason. It welcomes philosophers, scientists, and poets, under the auspices of Apollo and the Muses. The last planet, Saturn, is depicted as a pastoral paradise, a place where simplicity, temperance, and Nature are the main virtues (a portrait not unrelated to the theories of Jean-Jacques Rousseau (1712–1778)).

Finally, Zachiel reveals that Monime is actually the cousin, not the sister, of Céton as well as the heir of Georgia. They come back to Earth to help Monime to claim the throne, which she eventually acquires after a battle. She marries Céton and elevates him from consort to king status. Meanwhile he has found his lost father and the reunited family live happily ever after.

## 2.3 Reception and Inspiration

This long (900 pages in the 1787 edition!) novel was the third written by de Roumier-Robert. Since its publication, it has received mixed reviews. In 1765, when only the first two parts had been published, it was mocked as one of the worst novels inspired by *Gulliver's Travels* of Jonathan Swift[3] (1667–1745)—see Collective (1878: 185–186). Her other novels also received a similarly bad reception (Harth, 1992), although the lack of enthusiasm was not general. In his biography, La Porte and La Croix (1769: 80) recalled her "… fertile imagination …", and described de Roumier-Robert's writing as possessing

> … a striking mind and a sentimental tone that give to her novels a strong interest; her style is simple and natural; she does not adorn it with fashionable 'bel esprit'.

In their detailed and appreciative description of the *Voyages*, they mentioned that the text presents "… interesting episodes, naturally brought and chained together with art." (La Porte and La Croix, 1769: 99). In addition, twenty years later, Charles-Georges-Thomas Garnier considered the *Voyages* to be worthy ("… estimable …", see the editor's preface of de Roumier-Robert 1787a) and re-edited two of her novels, *Voages* and *Les Ondins*—those versions are today easier to find than the original leaflets.

Furthermore, in his *Mondes Imaginaires et Mondes Réels* (Flammarion, 1865), the astronomer and famous popularizer Camille Flammarion (1842–1925) devoted no less than 15 pages to the *Voyages* in Chapter XI of the second part of his book. He stated that such a long description was warranted because the book is the prototype of one kind of fiction (Flammarion, 1865: 512). He also stated that the "… feminine author of these travels possessed some skill in the mechanism of her novels." (Flammarion, 1865: 505), thereby showing his appreciation of the writings.

A more recent assessment of the work provides a balanced view, which summarizes well the weaknesses and strengths of the novel (Chambrionne, 2015):

> It has to be admitted that Voyages de Milord Céton is not the most readable of texts even by the standards of its day; it is not only prolix, repetitive, and inconsistent, but in some respects remarkably lacking in seemingly necessary intelligence although certainly not in boldness … Madame Robert's narrative also has some striking compensatory virtues. Not the least of those compensations is the sheer bizarrerie of certain parts of the narrative … Although undoubtedly long-winded and sprawling— in a manner far from unfashionable in its day—Voyages de Milord Céton's panoramic view of human life, divided up in order to emphasize different features in turn, does add up, jigsaw-style, to an original and worthwhile whole.

The novel possesses several sources of inspiration, the most obvious one of which is of course Fontenelle (1686), probably because of his relationships with the author's family. The idea of a plurality of worlds is indeed central to this travel throughout the Solar System, since each planet displays welcoming weather conditions and similarly hosts plants, animals, and human inhabitants. On a more specific note, one may find some direct borrowing from Fontenelle's text. For example, when discussing the origin of astronomy and geometry, pages 19–20 of de Roumier-Robert (1787a) uses but actually shortens the text on pages 28–31 of Fontenelle (1686). But that is not the only inspiring text used by de Roumier-Robert. The mentoring by the genie clearly recalls the novel *Les Avantures de Télémaque, Fils d'Ulysse* by François Fénelon[4] (1651–1715), as recognized by La Porte and La Croix (1769: 95). The detailed planetary descriptions, without a satirical tone, have links to the visit to Mercury described in the 1750 novel *Relation du Monde de Mer-* cure,[5] while any French travel to the Moon pays some dues to the pioneering novel by Cyrano (1657). However, C.-G.-T. Garnier refutes a direct linkage with Cyrano in his editor's preface of de Roumier-Robert (1787a) because marvelous images and caricatures are absent, being rather replaced by a "… fine and delicate criticism …" Finally, Chambrionne (2015) mentions a closeness to the Moon depicted in *Zamar …* by Tiphaigne de la Roche (1722–1774; 1754) and in *Le Voyageur Philosophe* … by Mr de Listonai (1761), both of which were published before *Voyages de Milord Céton …*

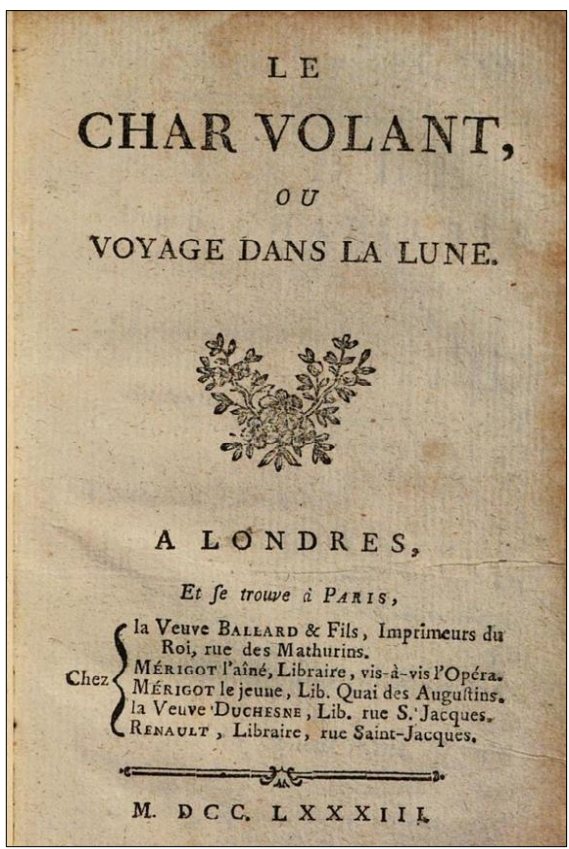

Figure 3: Front page of *Le Char Volant* by Wouters (1783, from Google books, id -FdU9nBjLVoC).

## 3　*LE CHAR VOLANT*

### 3.1　The Author

Cornélie Wouters was born in Brussels, probably in 1737 (Klaus, 2004). She married the Baron of Wasse (sometimes also written as Vasse), and travelled around Europe with him until his death (Collective, 1827: 256). She then settled in Paris, where this cosmopolitan woman hosted lively salons (Gevrey, 2020). She then had a short but intensive literary career. With her sister Marie, she produced several translations from English works, notably several theater plays and a history of famous men ("*Le Plutarque Anglois*"). Her translated plays contain a few adaptations to suit the taste for more formal language of the French public: for example, she softened rude language or direct sexual allusions (Klaus, 2007a). She also left some manuscripts popularizing natural sciences or chemistry (Collective, 1827: 257) although they seem lost today. In addition, she wrote fiction, with two libertinism novels and two travel novels, including a travel to the Moon in *Le Char Volant*, published in 1783 (Figure 3). Due to the French revolution, she lost access to her properties in England and Germany, leading her into some financial difficulties. She continued, however, to be involved in the intense discussions of the time. In 1790, she notably sent a text to the National Constituent Assembly to support citizenship access for the Jews (Klaus, 2007b). In an original move, clearly different from other contributors to that question, she did not try to excuse the perceived Jewish flaws (like usury) or to promise future conversions to Christianity if citizenship was granted. Rather, she referred to the first article of the 1789 *Declarations of the Rights of Man and the Citizen*—men are born and remain free and equal in rights—and asked for its simple, direct application. She added another, down-to-earth, argument playing on French pride: if France does not accept Jews as full citizens now, England (the hereditary enemy of France …) will do it soon, gathering all benefits of such a decision since Jews of all Europe would then move there, with their for-tune, rather than to France. Most probably, this contribution takes place in a grander scheme aiming at a recognition of all oppressed groups, with the possible ultimate goal to grant *women* the full benefits of citizenship … although that had to wait another century and a half. In the Revolution's aftermath, she had to leave Paris for a while, moving back to Brussels, but she could soon settle back in Paris, where she died in 1802.

## 3.2 The Novel

Eraste, a philosopher (one would say 'scientist' today), discovers a means to go to the Moon but he does not want to travel alone. He thus publishes a call for volunteers, and receives a number of responses. The examination of the candidates, with numerous humorous sides, constitutes the first part of the novel. Most of them want to go to the Moon for bad reasons, such as making money or gaining some advantage. In the end, Eraste accepts for the journey five people: an English scholar with his servant, wishing to improve astronomical knowledge, a gentleman who has lost his money, the widow from an unfaithful husband, and her son conceived with a malevolent lover.

The second part of the novel is a flashback: Eraste, coming back to Paris, tells the story of his lunar expedition. The Moon appears to be an ideal world with a single common language but a division into five kingdoms (or actually queendoms since the state heads are always female). Each kingdom is linked to a quality: love, fortune, justice, fame, and moderation. In each one, there is a magical place showing the excesses of those qualities—unbridled passion and jealousy in the love kingdom, for example—as they exist on Earth. There is an educational value in those theaters, as seeing the terrestrial defaults actually protects the lunar inhabitants from falling into the same traps as on Earth. Each traveller finally decides to stay in a different kingdom (although the two Englishmen stay together) but, under the suggestion of his queen, Eraste decides to go further, to the newly discovered Uranus. He however first needs to go back to Earth to get some needed material for his spacecraft, hence his return to Paris.

## 3.3 Reception and Inspiration

In her time, Wouters received a lot of praise for her works, which often went into several re-editions and foreign translations. She notably received two Gold Medals from the Swedish King for her *Plutarque Anglois*, which had been dedicated to him. *Le Char Volant* is no exception. One review of this work began with

> This work is by a foreign author, a woman; it thus doubly appeals the reader's indulgence, and one could claim it if the work needed this double recommendation. (Anonymous, 1784: 210–211).

Contrary to de Roumier-Robert, Wouters is included in the *Biographie Universelle* edited by Michaud (Collective, 1827: 256) with the flattering description that she is

> … endowed with a high character … learned without pedantry, kind without ambition to please; she spread in conversation the charms of a varied instruction, a gentle and playful philosophy, and an exquisite sensitivity.

Her renown was however attacked in the nineteenth century when one of her libertinism books was censored: it was considered too scandalous (Klaus, 2004). Her defense by a librarian (Lacroix, 1863) was not sufficient to keep her reputation as a writer. This probably explains why, contrary to de Roumier-Robert, she does not appear in Flammarion (1865). Her work nearly fully fell into oblivion during the twentieth century (Berditchevsky, 1990: 18; see, however, a mention of *Char Volant* in the bibliography of von Braun and Ordway, 1966: 227). However, there seems to have been a resurgence of interest in her work in recent years (Gevrey, 2020; Klaus, 2004; 2007a; 2007b).

The inspirations for Wouters' work probably are similar to those of de Roumier-Robert, as there were not many new works in the two decades between their novels. The description of the departure, however, with "… the chariot [rising] gently in the air …" in front of a huge crowd (Wouters, 1783: 115) recalls the Montgolfier balloon experiments invented the year before and whose first official flights took place in Paris the same year.

Regarding (old) lunar travels, Desile (2017) suggests two traditions (which may be also found in other voyage novels). On the one hand, placing the action on the Moon can be a clever stratagem to indirectly criticize the power or society of the time. Along with a common exaggeration, such novels often possess a satirical side, facilitated by the chosen location. Indeed, the Moon is associated with lunatics and madness, hence it is able to trigger quite colorful events (see e.g. Cyrano, 1657). On the other hand, an exotic place may rather be used to show what escapes us: an ideal society and all its benefits. Authors here play on perfecting humanity or show a pre-Fall, Eden-type society. One such case is Godwin (1648): in this novel, his proud and highly self-esteemed Spanish traveler meets tall lunar inhabitants who do not lie, murder, or experience sickness, and all speak the same language (as in *Le Char Vol-* *ant*). Mixed with those two main flavors, one also finds the unavoidable adventurous side of such faraway, hence dangerous, expeditions. In this context, both novels examined here play on both sides. There is a direct criticism of the contemporaries (the descriptions of the societies in five planets of *Voyages de Milord Céton*, the portraits of travel candidates in *Le Char Volant*), but without the actual satire found e.g. in Cyrano (1657). In parallel, both novels actually aim at elevated ideals, as in Godwin (1648). They hope to educate (especially *Voyages de Milord Céton*) or inspire, and push for these ideas to be actually put into practice on Earth (more on this subject will be discussed in Section 4.2).

## 4    SCIENTIFIC AND PHILOSOPHICAL CONTENT

Now that the two books and their authors have been presented, the main themes for such science-fiction stories are now examined in some detail: the astronomical content and the philosophical content.

### 4.1   Astronomy

In many science-fiction books, scientific discoveries and technologies provide a (credible) background or a means, not an aim. This is also the case of our two novels hence the majority of their text is not concerned with science itself. After all, they are not science popularization books, nor even belong to the 'hard science fiction' genre, replete with scientific information, of which Verne certainly is a prime representative. However, this does not mean that science is totally absent and, since the novels deal with the 'planets', the astronomical information mentioned in both books constitutes the main science content.

Being quite short (around 200 pages), *Le Char Volant* provides only a few such instances. Towards the end of chapter VII (Wouters, 1783: 199−200), the English scholar and his friend use their astronomical instruments. They observe the Earth, then Saturn and Jupiter. Five moons are said to be associated with Saturn: this was indeed the number known at the time of the novel's writing. Historically, Titan was discovered by Christiaan Huygens (1629−1695) in 1655, and Iapetus, Rhea, Tethys, Dione were found by Giovanni Domenico Cassini (1625−1712) about twenty years later. The next discoveries, those of Enceladus and Mi-mas, would occur six years after the novel's publication, in 1789, thanks to William Herschel (1738−1822).

The ring of Saturn is also mentioned as a "… cercle luminaire …" (which may be trans-lated as "enlighten circle") composed of many stars as is the Milky Way. This text clearly alludes to the fact that the ring(s) are made of numerous pieces, hence are not continuous ring(s). However, the impossibility for Saturn's ring(s) to be solid was only fully demonstrated by Pierre-Simon Laplace (1749−1827) (see Laplace, 1787). Since *Le Char Volant* was published four years before, it might be assumed that the author had discussed this question with Parisian astronomers. These would probably have been informal discussions, just as Verne often had in the next century. This would not be surprising since the Baroness is known to have participated in and even organized philosophical discussions (Gevrey, 2020).

Next, the text mentions Jupiter being enveloped by a bright cloud whose darkening enabled to see 4th magnitude stars with a Dollond telescope. The references to contemporary science are here much more obscure.

Finally, on the advice of one of the queens, Eraste decides to visit the new planet—Uranus' discovery had been announced only two years before, and its mention shows that the author kept up with the news (which is not too difficult in this case since that discovery had made headlines all over Europe and was certainly mentioned in Parisian salons).

Besides the peculiar Jupiter description, there are also a few other instances of less satisfactory sentences. At the beginning of Chapter V (Wouters, 1783: 170−171), the sky observed from the Moon is declared to have a "… totally different position …" than ours and the two Englishmen then discover stars unknown from Parisian astronomers in the direction of Mercury and Venus. However, the Moon is quite close to Earth compared to the stars, hence the celestial sphere is similar if not identical if seen from our satellite. Furthermore, after a few subsequent observations, the Englishmen finally decide that man does not know anything and that it is therefore best to admire the Universe and adore its creator than to try unveiling its secrets. They destroy their instruments, stay away from any science endeavor, and live happily afterwards. This awkward anti-science move strongly contrasts with the last event in the book, the hero's decision to pursue the adventure by traveling to Uranus.

Being much longer, the *Voyages de Milord Céton* provides more opportunities to discuss celestial information. However, most of it is of very limited quality. First, there are repeated allusions to astrology. Even the title and the overall plot are directly linked to old astronomy and astrology. Indeed, the novel is a travel to "… seven planets …" including the Moon and the Sun, whose status had changed when the heliocentric model was adopted in the previous century. Furthermore, the flavor of each planet is directly taken from basic mythological symbolism, e.g. money for Mercury, love for Venus, war for Mars. Moreover, there are repeated instances of astronomers actually conducting astrological studies and of constellations that influence fates or announce events, see notably de Roumier-Robert (1787a: 102, 270, 410, 503; 1787b: 37, 86). On the latter page, Nostradamus (1503–1566) is even said to be one of the best astronomers ever born on Earth! However, it must also be noted that astrology is mentioned to be "… a vain science …" whose predictions rest on the observation of human behavior (de Roumier-Robert, 1787a: 271).

Second, one can also find plain errors in the chapter about Mars: the Martian night appears illuminated by the Moon and the planet is said to be "… closer to the Sun than others …" (de Roumier-Robert, 1787a: 415 and 458, respectively). In the chapter on the Sun, one can also find a contradiction: it is clearly said that there is no night on this bright object, but then the text mentions celestial observations made from the Sun by astronomers (de Roumier-Robert, 1787b 19 vs. 37, 41–44, and 115). Finally, the attractive force on matter is described as proportional to mass and inversely proportional to the quarter of the distance (de Roumier-Robert, 1787b: 12–13), while Newton's gravitation law was well known—and used—at the time.

The only place with some correct astronomy is Chapter IV of the Fifth Chapter, the one about the Sun. Our star welcomes bright minds, which of course include the most famous astronomers (e.g. Aristarchus (310–210 BCE), Hipparchus (190–120 BCE), Ptolemy (100–170), Copernicus (1473–1543), Galileo (1564–1642), Brahe (1546–1601), Kepler (1571–1630), Gassendi (1592–1655), and Cassini).

On pages 41 to 44 of de Roumier-Robert (1787b), Céton, the narrator, begins to observe with one of them (unnamed but it can be reconstructed to be Edmund Halley, 1656–1742, see Figure 1). The astronomer talks with him about five variable stars and six nebulae, and then answers his question about the nature of comets. The text about stars and nebulae is directly inspired by Halley (1714a; 1714b).

From the list of variables from Halley (1714a), one object is missing, nova 1670 Vulpeculae. However, the five other variables are described using Halley's text, avoiding technical details such as names or coordinates. They appear in the following order: SN1572 (Tycho's supernova), Mira Ceti, P Cyg, SN1604 (Kepler's supernova) and χ Cygni. While the period found by Gottfried Kirch (1639–1710) for the latter star is correctly quoted, Mira Ceti is mentioned as making "… its revolutions in six years …" but that is probably an unfortunate typo (in French "ses révolutions en six ans" rather than the correct "sept révolutions en six ans"). Besides, only SN1604 is said to be of a different type, while Halley (1714a) wrote that for both supernovae. Strangely, Halley (ibid.) did not list two other variable stars then known, Algol and R Hydrae (their variabilities were reported in 1669 and 1704, respectively). Hence those stars do not appear in de Roumier-Robert's book. Adding the three missing objects to the five cited ones, the list appears complete for the time, as there was no further discovery before R Leonis in 1782.

The text related to nebulae uses longer excerpts of Halley (1714b). It is truly a direct translation of Halley's words and this helps identifying the unnamed astronomer showing the sky to Céton. Indeed, that astronomer says that he discovered the 4[th] nebula (actually the star cluster ω Cen) while working on a catalog of southern stars. The same sentence is found in Halley (1714b), the active voice ("I discovered …") of the novel replacing the passive voice ("… discovered by Mr Edmund Halley …") of the original text. Halley's list contains six nebulae whose description is copied in the novel (M42, M31, M22, ω Cen, M11, and M13 by order of appearance in the novel). However, this list was already outdated at the time of de Roumier-Robert's writing. Indeed, while the catalog of Charles Messier (1730–1817) was not yet available, two other references in French existed. The Swiss Jean-Philippe Loys de Chesaux (1718–1751) listed 20 nebulae in 1746 (Jones, 1969), while the French Nicolas-Louis de La Caille (1713–1762) reported 42 cases a few years later (La Caille, 1755). The very short list of de Roumier-Robert is probably explained by her seclusion: contrary to the Baroness, she was not directly engaged in the philosophical discussions of her age taking place in the Parisian salons hence she only used older references.

As the plot develops, it must be stressed that all planets in this novel are inhabited. This corresponds to the philosophy of the time, which had percolated through the upper society in France thanks to the famous 1686 book by Fontenelle on the plurality of worlds. While de Roumier-Robert had clearly read it (see above), she deviates from her model on two points. First, her planets seem to possess similar climates, whereas Fontenelle (1686) clearly explains several times that closeness to the Sun will influence planetary temperatures.

Second, she does populate the Sun, which Fontenelle presented as a hot bright world different from the planets (Fontenelle, 1686, chapter entitled "4th Night"). However, she is not alone in this interpretation: William Herschel is well known to have advocated the presence of solar residents.

Closer to Earth, the Moon is also inhabited in both novels. Lunar habitability had been a subject of debate for several years and it would remain for quite some time afterwards. Indeed, while Huygens avoided lunar inhabitants in his *Cosmotheoros*, Fontenelle populates the Moon without much hesitation and Verne (1869) still made a (cautious) allusion to a welcoming far side of the Moon in his novel *Autour de la Lune*. The choice of de Roumier-Robert and Wouters for the Moon is thus unsurprising and not at odds with their contemporary science.

Finally, it is also interesting to note that, like Fontenelle, de Roumier-Robert does not limit life to the Solar System. From the start, her genie repeats to the two orphans the famous quote by Metrodorus of Chios (fourth century BCE) on plurality: "… it would be as ridiculous to think that there is only one ear of wheat in a field as to consider a single world in the universe." (de Roumier-Robert 1787a: 27–28). Then, in the Jupiter chapter, she introduces a second genie who had traveled to these other worlds associated with the 'fixed stars'. Various species inhabit these worlds, but all are much less evolved than the creatures of our Solar System hence the genie does not recommend the two orphans to visit them, and so further travel is not envisaged …

## 4.2 Ideal Societies

Most science-fiction stories provide something more than just descriptions of travels or of exotic wildlife. They also discuss promises, defaults, and advantages of technologies, society organizations, and/or power systems. They can envisage contemporary existing cases, often pushed to their extremes to convey the point more easily, or present (good or bad) alternatives to current order. The two novels presented here are no exceptions in this respect. In particular, they both reveal interesting ideas concerning the place of women—the fact that their authors are female is certainly not unrelated to the choice of this subject.

In the eighteenth century, governing women were not just mere mirages. While several queens had already reigned in England in the preceding centuries, the Enlightenment era saw Maria Theresa (1717–1780) wisely ruling the Habsburg lands from 1740 to 1780 and Catherine the Great (1729–1796) as Empress of Russia from 1762 to 1796. There was no female ruler in France, and there still has not been one since then. However, there had been a few female regents (notably Blanche of Castille (1188–1252), Catherine de' Medici (1519–1589), Anne of Austria (1601–1666)), and literate persons certainly knew of the foreign examples. Around 1780, there were even several French fictions, written by men, envisaging female rulers, although the stories do not always end in a very happy manner for them (Grieder, 1989). It may thus not be totally surprising to find queens in the two examined novels, though they do present some interesting twists.

In *Le Char Volant*, the Moon is divided into five parts and all of them are ruled by queens. This is no happy accident as queens are succeeded by queens. There is no word about the possibility of a king in the novel and lunar men appear as very happy with the situation. Moreover, the spirits of the deceased queens are the true sources of inspiration for artists and scientists on Earth. The queens thus not only reign on the Moon, but are also providing the best realizations of the human mind on Earth. In this full realization of 'girl power', it thus came as no surprise that it is one of the queens who suggests to the hero to further explore the Universe.

In *Voyages de Milord Céton*, only one planet is ruled by women at all political levels: Venus. This is more a mythical need than a feminist advocacy, of course. However, two female heirs should be singled out in the story. On Mars, a fellow soldier of Céton, brave and of noble character, is finally revealed as a Martian princess who had to fight for her life, quite literally in her case. Moreover, the actual hero of the novel is not its narrator, Céton, but his 'sister' Monime. She is the one asking to go to space, fraternizing with local people, or making the most impressive appearances in the planetary courts—in other words, she makes most decisions. In the end, she is also the only one with the right to a throne and she fights to recover her land, a battle not done from a comfortable rear camp as she leads her army, sending out orders from the forefront and plunging into the melee to galvanize her troops. In this context, it may be interesting to note that cross-dressing (for both Monime, see Figure 2, and the Martian princess) and even a sex change (as Monime envisages on a comet used to go to Mercury) seem to be allowed.

In addition, equality constitutes the ideal foundation for any state in both novels. In *Le Char Volant*, the only aim of the five queens is making all citizens happy, and their agreement on this goal and how to reach it imply a "… perfect equality of conditions …" (Wouters, 1783:

132−133).  A similar equality, with promotion based only on merit, is evoked as an ideal throughout *Voyages de Milord Céton*.  It only truly exists in the perfect places of the Sun and Saturn.  On Saturn, for example, education is provided to all children and farmers receive the greatest respect as they are the ones providing the food so necessary to human life.  In contrast, travels to the Moon, Mercury, and Jupiter are the occasions to criticize court's plots and the undue sense of superiority by noble and/or rich people.  One may read for example (de Roumier-Robert 1787a: 183):

> Since the great lords can only become rich at the expense of the people, they try to persuade them that spirit, courage, feelings, kindness of heart, purity of language and great knowledge are innate in people of condition and that it is up to them alone to benefit from the sufferings and work of the poor.

The injustices created by such inequality are further personified by examples of poor but deserving humans, to emphasize the need for equality and merit recognition.  However, this equality is not limited to the usual economic status (noble or peasant, poor or rich), as often discussed during the Enlightenment.  It is also enlarged to an equality between genders.  The importance of this gender equality is particularly highlighted for education.  In the *Voyages de Milord Céton*, the Lunar Academy of Women is only occupied with defining fashions, but that is only a criticism of the female contemporaries of de Roumier-Robert and their education.  The ideal world, represented by the Sun, provides a very different view:

> In this world, men have no superiority over women unless virtue, science, common sense and reason give it to them.  It is certain that a woman can also possess all these gifts, especially when she receives the same education. (de Roumier-Robert, 1787b: 49).

Furthermore, the Sun is presented as the host of all great minds and that of course includes women,[6] like Emilie du Châtelet (de Roumier-Robert, 1787b: 73−74).  One may regret that the author lists many more famous men than illustrious women, in particular there are very few women of science, but that could be expected at a time when education and status were so vastly different that only extremely rare women could perform scientific research or follow artistic endeavors.  To underline this, one can also find in the Venus chapter a clear portrait of the women's status at the time (de Roumier-Robert, 1787a: 291−293):

> … in our world, men believe they have the right to demand anything.  They push their kindness to the point of attributing to women much weakness and more liveliness in their passions, and at the same time ask them for more strength than they themselves have to summon: I would like to ask them where does this exclusive privilege come from to be able to prevent all their desires, to yield to all their movements and to listen only to the voice of nature, while they hardly allow women the option of vegetating; they only see them as automatons who are only to be used as ornaments in a living room which they would like to decorate with various changes … women are constantly being shouted at, accused of inconsistency and infidelity, asked for unfailing virtue, and the unjust men who made the laws want to reduce them to hard slavery while they grant themselves full freedom ....

That text however contains another spike towards contemporary women, along with a path for the future:

> I am always amazed that women have not yet united … but so far they have been too coquettish and too dissipated to take seriously the interests of their sex.

One should remember that the narrator, the 'I', in those texts is supposed to a man, Céton, but it is clearly the voice of the author that is heard here.  The Venusian adventure provides a further demonstration by reversing the usual attributes of weakness and strength.  Indeed, when the orphans arrive on the planet, Céton is not allowed to take back his human form:

> … it is cautious not to expose [Céton] to temptations which it is nearly impossible for man to resist … Would it be possible that this [female] gender, which seems so delicate and weak to us, could nevertheless retain more firmness on occasions? (de Roumier-Robert, 1787a: 283).

Céton then remains a sort of fly while his 'sister' takes full part to the Venusian life.

In another quite modern move, both novels describe divorce as possible and even unproblematic.  In *Le Char Volant*, the wedding is said to be a free union, which can be broken at any time, by any of the spouses, without blame.  However, divorce is actually rare because the wise lunar people are always careful in their choices (Wouters, 1783: 191−192).  In *Voyages de Milord Céton*, divorce appears to be always granted on the perfect planet of Saturn while it is recommended that "… each spouse obeys and each one rules …" to have a happy wedding (de Roumier-Robert, 1787b: 336).

A last word may be said about religion.  It is difficult to find the usual marks of Christianity on the Moon, Sun, or planets of both novels.  Their extraterrestrials are not atheist, but the worship is more concerned by a supreme being like the one referred to by e.g. Voltaire. For example, in the *Voyages de Milord Céton*, there are temples to Apollo on the Sun, but the philosophy of the inhabitants is that "... the whole universe must serve [the divinity] as a temple and altar." (de Roumier-Robert, 1787b: 96). Besides, the mention to relieving poor ones only by "… blessings." (de Roumier-Robert 1787a: 153) also constitutes a criticism of the usual Christian practices of the time. This is again quite typical of Enlightenment philosophers.

## 5  CONCLUSION

In the France of the eighteenth century, two women proposed fictional interplanetary journeys: Marie-Anne de Roumier-Robert and Cornélie Wouters de Wasse. It is difficult to find two women more different than these two. The former had not traveled and lived a secluded life, while the latter was a cosmopolitan person, fluent in several languages, and deeply involved in the discussions of her time. Their novels also appear at first dissimilar: *Voyages de Milord Céton* by de Roumier-Robert is a long text repeatedly alluding to simplified antique symbolism (seven planets, Venus = love …), while *Le Char Volant* by Wouters is a short novel of adventure with a more modern tone and even some humor.

However, important parallels can also be drawn between them, in particular regarding their ideal vision of some celestial societies. Their advocacy of equality (especially gender equality and the need for recognition based on merit) still finds echoes today. Furthermore, while not exempt of scientific errors, both works used up-to-date astronomical information on little popularized subjects (the fragmented nature of Saturn's rings, the various properties of variable stars).

Unfortunately, the two novels and their authors are seldom mentioned in astronomical circles or in recent works on literature or the history of science fiction and proto-science fiction. For example, one of the longest recent mentions consists of a few lines on *Voyages de Milord Céton* and one sentence on *Char Volant* (Roberts, 2006: 79 and 81, respectively). Nevertheless, the portrait of the society of their time along with their dreamed ideals (some still not fulfilled today!) makes these books worth reading, even if their astronomical content is scarce.

## 6  NOTES

1.  Fontenelle first wrote a few theater plays but, with their lack of success, he turned to science popularization. He became famous in 1686 with *Entretiens sur la Pluralité des Mondes*, which presents the contemp-orary knowledge of the Solar System (heliocentrism, existence of other moons) and discusses the possible presence of inhabitants on celestial objects. The pleasant text went not only through several re-editions in France, but it was also translated in several languages abroad, becoming a reference on astronomy in the early Enlightenment period. With his long life, the jovial Fontenelle was also an unavoidable landmark in the literary microcosm of the time.

2.  The Scottish economist John Law (1671–1729) served as financial advisor in France in the early years of the reign of King Louis XV, rising to the position of Controller General of Finances in 1720. Pushing for the use of paper money, he created the first bank of the country, where people could exchange gold and silver against banknotes. This move was seen as a solution to the huge debt of the country but it became problematic when too many banknotes were printed by the bank and speculation occurred over shares of the Mississippi Company linked to it. The bubble exploded in mid-1720 when people asked to convert notes back into coins.

3.  The Anglo-Irish cleric Jonathan Swift wrote poetry, political pamphlets, and satires. He is most known for this last kind of writing, and *Gulliver's Travels* is considered as his masterpiece. As was the case for Cyrano, Swift was inspired for this work by *The Man in the Moone* by Francis Godwin (1648), who actually was his relative.

4.  François Fénelon was born in a poor noble family but he could get a philosophy degree from the University of Cahors (southwest France). He then started an ecclesiastic career and became a prominent man in the French court. Fénelon wrote several books, notably *Les Avantures de Télémaque, Fils d'Ulysse* (1717). This text was one amongst those aimed at the heir of the throne, of whom he was the tutor, but it was leaked by a servant and began to circulate broadly. In it, a mentor demonstrates the shortcomings of antique states to a pupil, which was an indirect criticism of Louis XIV politics and it provoked his disgrace. Note that the mentor is called in this book "Mentor", while the subtitle of *Voyages de Milord Céton* is *Le Nouveau Mentor* (i.e., the new mentor), most probably indicating the direct filiation between the works, which share the same spirit (i.e., educating by showing bad behaviour in an imaginary context).

5.  The front page of *Relation du Monde de Mercure* does not quote any author, and this book is said to be anonymous both by

Flammarion (1865: 472) and in the editor's preface of the reedition by C.G.T. Garnier in 1787 (it is included in Volume 16 of his *Voyages Imaginaires* collection of which de Roumier-Robert's novel occupies volumes 17 and 18). Today, the author is quoted as 'Chevalier de Béthune', notably in the Gallica database of the French National Library or in Roberts (2006: 78).

6. The full list of those "… talented and illustrious …" women mentioned in the text is: Madame de Maintenon (1635–1719), Marquise de Sévigné (1626–1696), Sappho (630–570 BCE), Deshoulières (either Antoinette, 1637–1694, or her daughter Antoinette-Thérèse, 1659–1718), Madame de Villedieu (1640–1683) and Emilie du Châtelet (1706–1749). The latter one is the only scientist in the list, all others belonging to the literatary world. The text mentioned "… several others …" without elaborating. Several other female writers, known in France at the time, could indeed have been mentioned, such as Mademoiselle de Scudéry (1607–1701), Margaret Cavendish (1623–1673) or Madame de La Fayette (1634–1693). Other female scientists were also known at the time too, notably linked to astronomy, e.g. Hypatia (~360–415), Elisabeth Catherina Hevelius (1647–1693), Maria Margaretha Kirch (1670–1720) or Nicole Reine Lepaute (1723–1788). In particular, the latter one participated in the Parisian calculations linked to the return of Halley's Comet, in 1759, only a few years before de Roumier-Robert's writing. The restricted list of de Roumier-Robert therefore appears somewhat surprising, but the reasons behind it remain unfortunately unknown.

## 7 ACKNOWLEDGMENTS

The author warmly thanks Francis Van Dam who brought Cornélie Wouters to her attention by sending the lively paper written by D. Berditchevsky: this triggered a curiosity that led to this contribution. The author also thanks Myron Smith for his language advice, Thomas Beyer for his careful reading, Sian Lucca for a suggestion, the FNRS for funding, as well as the Liège Library for its help. She acknowledges the use of ADS, CDS, Gallica, archive.org, and jstor databases. Note that all translations were made by the author. For Wouters' book see https://books.google.be/books?id=FdU9nBjLV oC and the Mercure de France's article can be found at https://books.google.be/books?id=idCB953Wj OYC; other ancient references quoted below (including de Roumier-Robert's novel) are available on https://gallica.bnf.fr

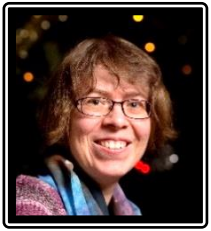

**Yaël Nazé** is a FNRS Senior Researcher working at the University of Liège, Belgium. She got her PhD in 2004 and became an IAU member in 2012.

Most of her research concerns the observational properties of massive stars, but she also performs some multi-disciplinary research on historical or sociological aspects of astronomy. In this context, she has notably published on Newton's chronology, astronomical knowledge of the Belgian public, and Max Ernst's work on Wilhelm Tempel. Her books include *Art & Astronomie* (2015, Omniscience) and *Astronomie de l'Étrange* (2021, Belin).